\documentclass{article}
\usepackage{spconf,amsmath,graphicx}

\usepackage{cite}
\usepackage{bm}
\usepackage{hyperref}
\usepackage{booktabs}
\usepackage{paralist}
\usepackage{float}
\usepackage{subcaption}
\usepackage{adjustbox}
\usepackage{fancyhdr}
\usepackage{multirow}
\usepackage{balance}
\usepackage[table]{xcolor}
\newcolumntype{P}[1]{>{\centering\arraybackslash}p{#1}}
\newcolumntype{R}[1]{>{\raggedleft\arraybackslash}p{#1}}

\newcommand{\sstitle}[1]{\smallskip\noindent\textbf{#1.\/}}

\newcommand{\eg}{e.\,g.,\ }
\newcommand{\ie}{i.\,e.,\ }

\ninept

\title{Knowledge Transfer For On-Device Speech Emotion Recognition \\ with Neural Structured Learning}

\name{Yi Chang${^1}$, Zhao Ren${^2}$, Thanh Tam Nguyen${^3}$, Kun Qian${^4}$, and Björn W. Schuller${^{1,5}}$ 
\thanks{%Technical Report.\\
This research was partially funded by the Federal Ministry of Education and Research (BMBF), Germany under the project LeibnizKILabor with grant No.\,01DD20003, and the research projects ``IIP-Ecosphere”, granted by the German Federal Ministry for Economics and Climate Action (BMWK) via funding code No.\,01MK20006A, the Ministry of Science and Technology of the People's Republic of China (No.\,2021ZD0201900), the National Natural Science Foundation of China (No.\,62272044), the National High-Level Young Talent Project, and the BIT Teli Young Fellow Program from the Beijing Institute of Technology, China.
}}
\address{$^1$GLAM – Group on Language, Audio, \& Music, Imperial College London, United Kingdom\\
$^2$L3S Research Center, Leibniz University Hannover, Germany\\
$^3$Griffith University, Australia\\
$^4$School of Medical Technology, Beijing Institute of Technology, China\\
$^5$Chair of Embedded Intelligence for Health Care and Wellbeing, University of Augsburg, Germany\\
}
\begin{document}
%\ninept
%
\maketitle
\begin{abstract} %BS: I corrected the abstract for language :)
Speech emotion recognition (SER) has been a popular research topic in human-computer interaction (HCI). As edge devices are rapidly springing up, applying SER to edge devices is promising for a huge number of HCI applications. Although deep learning has been investigated to improve the performance of SER by training complex models, the memory space and computational capability of edge devices represents a constraint for embedding deep learning models. We propose a neural structured learning (NSL) framework through building synthesized graphs. An SER model is trained on a source dataset and used to build graphs on a target dataset. A relatively lightweight model is then trained with the speech samples and graphs together as the input. Our experiments demonstrate that training a lightweight SER model on the target dataset with speech samples and graphs can not only produce small SER models, but also enhance the model performance 
%over models with speech samples only and those with classic transfer learning strategies.
compared to models with speech samples only and those using classic transfer learning strategies.

%edge device

%Neutral Structured Learning as a transfer learning strategy through building synthesized graphs

%DEMOS as the target dataset

%RAVDESS as the source dataset

\end{abstract}
\section{Introduction}
\label{sec:intro}

Speech emotion recognition (SER), which aims to recognise emotional states from speech, has been a popular research topic in the domain of human-computer interaction (HCI)~\cite{9383000}. SER has been applied to a range of applications, including call centres, education, mental health, computer games, and many others~\cite{AKCAY202056}. 
In particular, 
%as an important perception, 
speech signals provide rich and complementary information to other modalities, \eg images, biosignals, social media, etc~\cite{8805181}. SER can not only improve the performance of emotion recognition when combined with other modalities in a multimodal system, but also enable machines to perceive human emotions when other modalities are not available, such as in audio-only call centres~\cite{9383000}.

Deep learning has been widely applied to SER with many model architectures, including spectrum-based and end-to-end models. Spectrum-based models process spectrum features from speech signals~\cite{WU2011768, 10.1145/3267935.3267948}, while end-to-end models directly process raw speech signals~\cite{9413144, 8462677}. Convolutional neural networks (CNNs), recurrent neural networks (RNNs), and their variants (\eg transformers) have been commonly used to build either end-to-end or spectrum-based models for SER~\cite{9414314,ZHAO2019312}. 

Improving the performance of SER faces two challenges. First, % due to potentially biased and time-consuming annotation for emotional states on speech data, most emotional speech datasets are small-scale~\cite{8638999}. 
creating emotional speech datasets with high-quality annotations is a time-consuming and potentially biased process, leading to small-scale datasets~\cite{8638999}. Second, 
% as SER has the potential to be applied to edge devices for plenty of Internet-of-Things applications, 
with the increasing demand for SER in Internet-of-Things (IoT) applications, it is essential to train lightweight neural networks for efficient model development. However, directly fine-tuning complex SER models pre-trained on large-scale data places a high demand on computing systems~\cite{NEURIPS2020_eae15aab}. 

More recently, neural structured learning (NSL) was proposed to add structured signals (\eg graphs) as the model input in addition to the original data~\cite{nsl}. In particular, NSL was developed to solve the data labelling problem of semi-supervised learning, and to build adversarial training 
%for overcoming the model security problem caused by 
against adversarial attacks~\cite{nsl}. Inspired by NSL, it is promising to construct a graph with a pre-trained model to break the bottleneck caused by small-scale labelled data and edge devices. 

In this study, we propose an NSL framework to transfer the knowledge of a large, pre-trained SER model to a smaller model with graph. To the best of the authors' knowledge, there were only few studies using NSL for SER~\cite{UDDIN2020103775}. The contributions of our work are twofold: i) transferring model knowledge through an NSL-generated graph can improve performance by leveraging multiple databases; ii) the proposed NSL framework can train lightweight neural networks without a high requirement of computing resources. Evaluated on an emotional speech dataset, our NSL framework outperforms models trained on the original data only and those trained with classic transfer learning strategies.

%The further      @Zhao, I have written something about NSL and wav2vec in the section methodology. Please check and feel free to re-organize sections, and delete unnecessary (sub)sections.

%Transfer Learning, edge device

%Structured learning

\sstitle{Related Works}
\emph{Deep Learning for SER.} 
%Compared with traditional machine learning models (\eg SVM, decision tree), CNNs and RNNs have been widely applied for SER because of their superior performance. 
As previously mentioned, deep learning is mainly applied for SER with spectrum-based and end-to-end models. %Due to smaller data size of extracted spectrums as compared to raw speech signals, spectrum-based models can be shallower and more efficient than an end-to-end model. 
Spectrum-based models are typically shallower and more efficient than end-to-end models due to smaller data size of extracted spectrums compared to raw speech signals.
%In~\cite{ZHAO2019312}, 2-D CNNs and long short-term memory (LSTM)-RNNs were combined to learn representations from log Mel spectrograms extracted from speech. 3-D CNNs were proposed in~\cite{8273628} to extract spectro-temporal features more effectively. 
On the other hand, end-to-end models save the time of selecting suitable spectrum types and can extract complementary features over fixed spectrums.  
An end-to-end model was developed based on 1D CNNs for learning spatial features in SER~\cite{MUSTAQEEM2021114177}. Additionally, stacked multiple transformer layers were utilised in~\cite{9414314} to better extract global feature dependencies from speech.
%An end-to-end SER model in~\cite{MUSTAQEEM2021114177} was developed based on 1-D CNN implementing the multi-learning strategy for spatial salient emotional features and contextual dependencies extraction in parallel. 
%Stacked multiple transformer layers were utilised in~\cite{9414314} to better extract global features from speech. 
More recently, wav2vec models, which include CNNs and transformers, were trained with self-supervised learning on unlabelled speech data~\cite{baevski2020wav2vec}. Wav2vec models have been applied to generate speech embeddings for SER~\cite{pepino21_interspeech, 9747095}.
%Pretrained wav2vec models are applied to generate embeddings for SER~\cite{pepino21_interspeech, 9747095}. 
%\zr{this paragragh is confusing. better to re-organise and consider moving some sentences of the 2nd paragragh to this part.}
%\yc{Differently, our work focus on lightweight CNNs that can be run on edge devices.}

\emph{Transfer Learning.} 
Transfer learning was designed to transfer knowledge learnt from a large-scale source dataset to a smaller target dataset for better performance in an efficient manner~\cite{transfer2}. Transfer learning has been applied to SER due to limited labelled emotional speech data. For instance, a source dataset with a large number of languages was utilised along with a small amount of target data for training deep belief networks in~\cite{tf1}. 
Various approaches to transfer learning aim to gap the data distribution difference between the source dataset and the target dataset~\cite{transfer2}: instance-based transfer learning achieves this by re-weighting samples in the source dataset during training; feature-based transfer learning attempts to learn new feature representations from original ones; parameter-based transfer learning targets to find shared model parameters between the source domain and the target domain; relational knowledge transfer learning works on relational domains with non-i.i.d.\ data and builds mapping of relational knowledge~\cite{transfer2}.
%In~\cite{10.1145/3462244.3481003}, a pretrained ResNet model for speaker recognition is leveraged to efficiently process the input speech with variable length. 
% Different from classic transfer learning strategies, which focus on fine-tuning the pre-trained model on the target dataset, our proposed work achieves knowledge transfer through the NSL framework. 
These transfer learning strategies mostly (instance-based, parameter-based, and relational ones) require training one or two models with the same architecture on both source and target data. The proposed knowledge transfer via the NSL framework is an extension of feature-based transfer learning. 
%We add graphs as additional inputs to the model on the target data rather to have more information than features learnt from extracted representations only.
Compared with the transferred representations, the additional graphs fed into the model on the target data have more regularised and complementary information.

%Instead of fine-tuning the pre-trained model on the target dataset, our proposed work achieves knowledge transfer through the NSL framework. 
%\zr{1. the use of this ref~\cite{10.1145/3462244.3481003} seems not useful.
%2. What are classic transfer learning strategies? You did not introduce. Re-write I think.}
%\yc{done}

\emph{Neural Structured Learning.}
Neutral Structured Learning (NSL) adds additional structured signals (\eg graghs) as a regularisation term to maintain the latent structural similarity among input signals. In~\cite{UDDIN2020103775}, Mel frequency cepstrum coefficients and NSL-generated structured signals were combined for SER. However, % transfer learning is not applied, all features are based on one single dataset, and structured signals are not clearly specified.
there was no knowledge transfer in that work and the type of structured signals were not clearly specified.
%\zr{1. adversarial training is training strategy, not structured signal
%2. I don't know why to mention transfer learning, is transfer learning available in that work?}
%\yc{1. NSL generates adversarial examples as the structured signals and adversarial examples are used to construct the graphs. 2. done}

\section{Methodology}
\label{sec:method}
\begin{figure}
    \centering
    \vspace{-20pt}
    \includegraphics[width=0.85\linewidth]{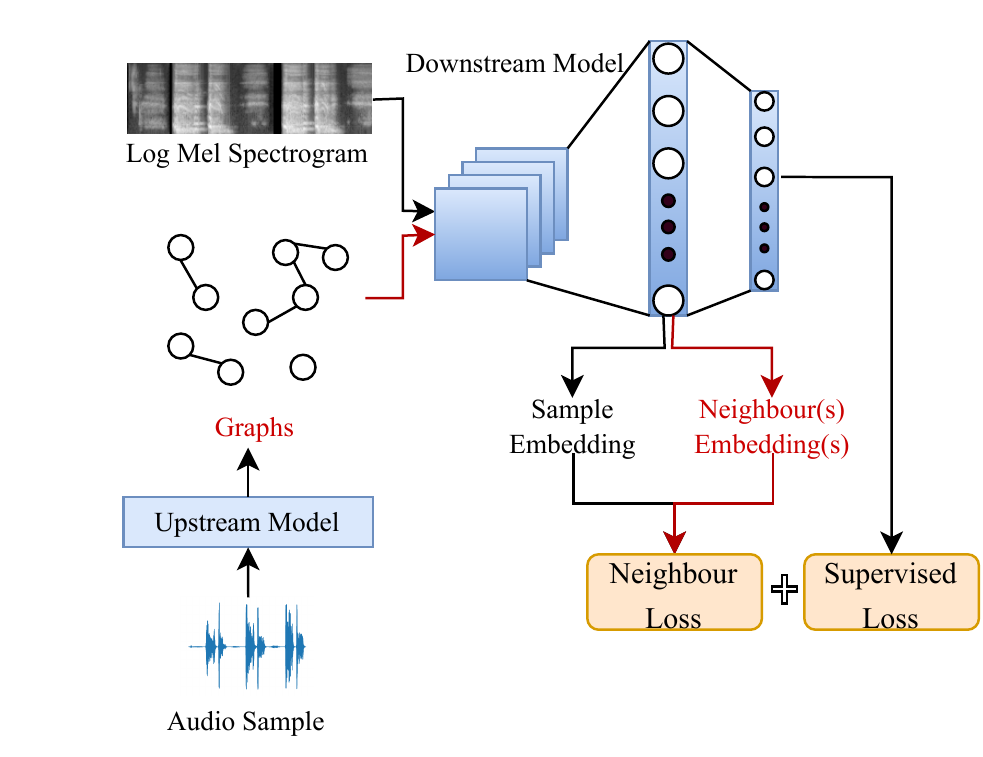}
    \caption{The framework of the applied NSL for SER. Graphs act as structured signals and, along with log Mel spectrograms, are fed into a downstream lightweight model. The final loss function is the weighted sum of neighbour loss and supervised loss.}
    \vspace{-15pt}
    \label{fig:framework}
\end{figure}

As depicted in Figure~\ref{fig:framework}, with the help of a model trained on a related emotional source speech dataset, 
%embeddings of audio samples in the target dataset are extracted and then graphs are constructed accordingly.
embeddings of audio samples in the target dataset are extracted and then used to construct graphs accordingly.
Log Mel spectrograms of the audio samples, along with the graphs, are fed into a relatively lightweight model. Afterwards, for each training sample, the loss function has two parts: 
%(1) the distance between its corresponding feature embedding and its neighbors' corresponding embeddings are reduced by the neighbor loss to maintain the structural similarity; (2) the supervised loss is calculated based on the predicted label and the true label in supervised learning. 
(1) the neighbor loss, which reduces the distance between the embedding of a sample and the embeddings of its neighbors to maintain structural similarity, and (2) the supervised loss, which is calculated based on the predicted label and the true label in supervised learning.

%The weighted sum of neighbor loss and supervised loss is the final loss function to be minimised during the lightweight model development. 
% \zr{
% 1. it is not clear to say "related source" - related to what?
% 2. current dataset -> the target dataset
% 3. fed into the lightweight models -> fed into a lightweight model
% 4. "of its own intermediate embedding as well as its neighbor(s) embedding(s) are reduced through the neighbor loss to maintain structural similarity" -> reedit
% 5. "predicted label " from data samples?}
% \yc{all done; related is a general idea and related to the target dataset?}

%We apply the NSL to transfer knowledge learnt from a pre-trained wav2vec 2.0 model on a reliably labelled dataset to lightweight models. The knowledge learnt on the source dataset is applied to construct graphs, which connect similar nodes in the target dataset. The distance between sample embedding and its neighbor(s) embeddings from a selected intermediate layer act as a regulisation loss during the current model training procedure. The applied NSL-based transfer learning framework is depicted in Fig~\ref{fig:framework}.

\subsection{Upstream and Downstream Models}
The upstream model denotes a large pre-trained model on a source dataset in SER, while the downstream model on the target data is lightweight and possibly run on edge devices. As noted in Section~\ref{sec:intro}, spectrum-based models can be smaller than end-to-end models due to different inputs. Therefore, the upstream model is an end-to-end model and the downstream one is a spectrum-based model herein.
%for the target dataset with log Mel spectrograms and graphs as input. 
%Moreover, compared with raw audio samples, the extracted log Mel spectrograms require less storage space. 

\emph{Upstream. }Due to its powerful capability for representations extraction, the upstream model utilised herein is wav2vec 2.0~\cite{baevski2020wav2vec}, which has been widely applied in 
%the speech recognition~\cite{9747379, NEURIPS2021_b17c0907, linke-etal-2022-conversational, cambara-etal-2022-recycle} and 
SER~\cite{pepino21_interspeech, 9747417, 9747870}. Wav2vec 2.0 is an end-to-end model composed of a CNN module as the latent speech \emph{feature encoder} and a \emph{Transformer module} for capturing global contextual dependencies. The output of the encoder module is discretised with product quantisation~\cite{5432202} for self-supervised training. The pre-trained wav2vec 2.0 model applied in this work is trained on the 960-hour Librispeech corpus~\cite{panayotov2015librispeech}. An audio classification head is then added with two linear layers %`FC1' and `FC2' for the task of SER. 
for the SER task. 
% With the added classification head, the pre-trained wav2vec 2.0 model is fine-tuned on the source dataset and afterwards it is the above mentioned upstream model.
After fine-tuning the pre-trained wav2vec 2.0 model on the source dataset, it becomes the aforementioned upstream model.

%and has an audio classification head.

\emph{Downstream. } Compared to raw audio signals, the extracted log Mel spectrograms require less storage space. Several representative lightweight downstream CNN models on the target dataset are selected: VGG-15, ResNet-9, and CNN-6, whose parameter numbers are 14.86\,M, 4.96\,M, and 4.44\,M, respectively. 
However, the wav2vec 2.0 model has over 90.37\,M parameters, making the above downstream models relatively lightweight.
% To this end, they are relatively lightweight.

\subsection{Neural Structured Learning}
%NSL was proposed in~\cite{nsl} to leverage structured signals to regularise the neutral network development. 
With graphs as the structured signals, NSL maintains the structural similarity between the speech samples. 
%In this way, several advantages can be obtained: higher accuracy, better robustness, and unlabelled data utilisation~\cite{nsl}. 
%The pre-trained wav2vec 2.0 model is first fine-tuned on the source dataset and then applied to generate the embeddings for audio samples in the target dataset.
In this way, NSL has the potential to improve the model performance with a low requirement of computational resource during training the downstream model.
\vspace{-5pt}
\subsubsection{Graph}
\vspace{-5pt}
The input to the downstream model includes data nodes $D$ and the graph $G$. Specifically, $D = \{(x_i, y_i)\}, i=1,..., N$, where $x_i$ is a data sample, $y_i$ is its emotional state, and $N$ is the number of training samples; The graph $G = (D, E)$, where $E$ denotes the bi-directional edges connecting similar nodes. A pair of connected nodes in $G$ is a neighbour of each other $\mathcal{N}=(x_i, x_j), i\neq j$. The neighbours of $x_i$ are computed by
\vspace{-5pt}
\begin{equation}
    \{x_k |S(h(x_i), h(x_k))>\epsilon,\ k=1, ..., N\ \mbox{and}\ k\neq i\},
    \vspace{-5pt}
\end{equation}
where $S$ is the cosine similarity, and $h$ is the feature embedding extracted by the upstream model.
%its graph $G = (X, E)$ is calculated, where $E$ denotes the bi-directional edges connecting similar nodes. 
%Connected nodes in $G$ are neighbors $\mathcal{N}$. To build $G$, for each $x_i$, with a pre-defined maximum number of neighbors $n$ and a similarity threshold $\epsilon$, the edge is constructed based on the output embeddings $\mathcal{H}$ from the upstream model. 
The graph $G$ with structured similarity information is then fed into the current lightweight downstream model to achieve the knowledge transfer.

 %When constructing the graph, we regard each audio sample in the target dataset for model development as a node; an edge is built if the connecting nodes' embeddings are similar enough (\eg cosine similarity). It's worthy noting that embeddings are output of the second last fully connected layer `projector' of the fine-tuned wav2vec 2.0 model. The structured similarity information reflected by the graph are fed into the model developing procedure through the neighbor loss.

\label{sec:nsl}
%BS: Please change in the image Neighbor --> Neighbour (2x!) Also, I suggest writing every word with initial capital (e.g., Audio sample --> Audio Sample)

\vspace{-5pt}
\subsubsection{Loss Function}
\vspace{-5pt}
The loss function applied in this work is the weighted sum of two components, supervised loss and neighbour loss, as follows:
\begin{equation}
\begin{aligned}
    \mathcal{L}=&\sum_{i = 1}^{N}\mathcal{L}_{\mbox{supervised}} (y_i, \hat{y}_i) + 
    \alpha \sum_{i = 1}^{N} \sum_{\{x_k\}} \mathcal{L}_{\mbox{neighbour}} (p(x_i), p(x_k)),
\end{aligned}
\label{eq:loss}
\end{equation}
where $\mathcal{L}_{\mbox{supervised}}$ is calculated by the cross entropy loss function for classification based on the target $y$ and predicted $\hat{y}$; $\mathcal{L}_{\mbox{neighbour}}$ reflects the distance between the embeddings $p(x_i)$ and $p(x_k)$ from the intermediate layers of the downstream model. 
%The distance function can be $L_1$, $L_2$, or $(1- \mbox{cosine}\ \mbox{similarity})$. 
We use $(1- \mbox{cosine}\ \mbox{similarity})$ as the distance function and $\alpha$ is a multiplier.
%for the weighted sum of the two loss functions. 
%based on the target hidden layer output embeddings $\mathcal{p}$ of sample $x_i$ and its neighbors $\mathcal{N}_{x_i}$; $\alpha$ is the multiplier for the weighted sum. 

By minimising the neighbour loss, the similarity among a sample and its neighbours is maintained. As a result, the transferred knowledge from the source dataset is used on the target dataset.

\section{Experiments}
\label{sec:exp}
\subsection{Databases}
\label{sec:data}

\sstitle{RAVDESS} The Ryerson Audio-Visual Database of Emotional Speech and Song (RAVDESS) is a multi-modal English dataset containing 1,440 speech recordings. It was recorded from 24 professional actors (12 females and 12 males) and labelled into eight emotional classes: \emph{angry},  \emph{calm}, \emph{disgusted}, \emph{fearful}, \emph{happy}, \emph{neutral},  \emph{sad}, \emph{surprised}. 
%BS: I sorted these by alphabet and turned them all into adjectives (disgust --> disgusted, surprise --> surprised)
% Each speaker has eight recordings for each emotional class, apart from \emph{neutral} including four samples.
Each emotional class has eight recordings from each speaker, except for \emph{neutral}, which has four.
We split the dataset in a speaker-independent and gender-balanced manner 
%and the details can be found 
shown in Table~\ref{tab:ravdess}.

Besides its wide applications in SER~\cite{AKCAY202056}, the reasons why we choose RAVDESS as the source dataset are that it is recorded from professional actors in a well-controlled environment, its emotional labels have high levels of validity, and the dataset is gender-balanced. 

\begin{table}[ht]
\centering
\caption{Split details of the source dataset RAVDESS.}
%\begin{tabular}{llllll}
\vspace{-10pt}
\begin{adjustbox}{width=0.7\columnwidth}
\begin{tabular}{l|R{.8cm}R{.8cm}R{.8cm}|R{.8cm}|r}
\toprule
\# & Train & Val & Test & $\bm{\sum}$ & F / M \\ \hline
\multicolumn{1}{l|}{Speaker} & 10 & 8 & 6 & 24 & 12 /\;\, 12 \\ \hline
$\bm{\sum}$ & 600 & 480 & 360 & 1,440 & 720 / 720 \\ 
\bottomrule
\end{tabular}
\vspace{-15pt}
\label{tab:ravdess}
\end{adjustbox}
\end{table}

\sstitle{DEMoS} The Database of Elicited Mood in Speech (DEMoS)~\cite{parada2019demos} is an Italian speech corpus with $7.7$ hours of audio samples recorded from $68$ speakers (45 males and 23 females). Besides the $332$ neutral speech recordings, there are in total $9,365$ audio samples annotated into seven classes: \emph{anger}, \emph{disgust}, \emph{fear}, \emph{guilt}, \emph{happiness}, \emph{sadness}, and \emph{surprise}. Similar to prior work~\cite{ren2020enhancing} on the DEMoS data, the minority \emph{neutral} class is not considered and the remaining $9,365$ emotional speech samples are divided into 40\,\% train, 30\,\% validation, and 30\,\% test sets with speaker-independent strategy. The detailed data distribution of the DEMoS is depicted in Table~\ref{tab:demos}.

\begin{table}[ht]
\centering
\caption{Emotion distribution of the target dataset DEMoS dataset.}
%\begin{tabular}{llllll}
\vspace{-10pt}
\begin{adjustbox}{width=0.7\columnwidth}
\begin{tabular}{l|R{.7cm}R{.7cm}R{.7cm}|R{.7cm}|r}
\toprule
\# & Train & Val & Test & $\bm{\sum}$ & F / M \\ \hline
\multicolumn{1}{l|}{Speaker} & 27 & 25 & 16 & 68 & 23 /\;\;\;\;\, 45 \\
\hline
\multicolumn{1}{l|}{Anger} & 586 & 531 & 360 & 1,477 & 400 /\;\;\, 729 \\
\multicolumn{1}{l|}{Disgust} & 666 & 608 & 404 & 1,678 & 596 / 1,082 \\
\multicolumn{1}{l|}{Fear} & 461 & 404 & 291 & 1,156 & 524 /\;\;\, 871 \\
\multicolumn{1}{l|}{Guilt} & 453 & 395 & 281 & 1,129 & 415 /\;\;\, 741 \\
\multicolumn{1}{l|}{Happiness} & 561 & 471 & 363 & 1,395 & 516 /\;\;\, 961 \\
\multicolumn{1}{l|}{Sadness} & 606 & 543 & 381 & 1,530 & 349 /\;\;\, 651 \\
\multicolumn{1}{l|}{Surprise} & 396 & 358 & 246 & 1,000 & 532 /\;\;\, 998 \\ \hline
$\bm{\sum}$ & 3,729 & 3,310 & 2,326 & 9,365 & 3,332 / 6,033 \\ 
\bottomrule
\end{tabular}
\vspace{-40pt}
\label{tab:demos}
\end{adjustbox}
\end{table}

%MSP-Podcast corpus~\cite{Lotfian_2019_3} is the one of the largest speech emotional dataset in the community, and it contains 86,389 crowd-sourced audio segments amounting to more than 100 hours. There are in total of eight emotional classes: \emph{anger}, \emph{happiness}, \emph{sadness}, \emph{disgust}, \emph{surprised}, \emph{fear}, \emph{contempt}, and \emph{neutral} and \emph{other}. Like prior works~\cite{}, only four classes are considered in this work, including \emph{anger}, \emph{happiness}, \emph{sadness}, and \emph{neutral}. We applied the officially provided train set, validation set, test set 1. The test set 1 is more gender balanced. The data distribution can be found in Table~\ref{tab:msp}. 

%\begin{table}[!h]
%    \centering
%    \footnotesize
%    \vspace{-5pt}
%    \caption{The data distribution of the MSP-Podcast database. }
%    \label{tab:msp}    
%    \vspace{-10pt}
%    %\scalebox{0.85}{
%    \begin{tabular}{l|p{1cm}p{1cm}p{1cm}p{1cm}}
%    \toprule
%         \#& \textbf{Train} & \textbf{Val} & \textbf{Test} & $\bm{\sum}$ \\
%         \hline
%         \textbf{Neutral}& 19\,563 &  2\,942 & \ \ 5\,362 & 27\,867\\
%         \textbf{Happiness}& 11\,740 &  1\,650 & \ \ 3\,851 & 17\,241 \\
%         \textbf{Anger} & \ \ 3\,486 & \ \ \,975 & \ \ \ \ \,775 & \ \ 5\,236 \\
%         \textbf{Sadness} & \ \ 2\,176 & \ \ \,346 & \ \ \ \ \,572 & \ \ 3\,094 \\
%         \hline
%         $\bm{\sum}$ & 36\,965 & 5\,913 & 10\,560 & 53\,438 \\
%         \bottomrule
%    \end{tabular}
%    %}
%    \vspace{-10pt}
%\end{table}

%\todo{TODO: It seems there is no description about the labels.}

\subsection{Experimental Settings}
\label{sec:expsetting}
\sstitle{Evaluations Metrics}
Similar to previous works on the DEMoS, the Unweighted Average Recall (UAR) is utilised as the standard evaluation metric to mitigate the class imbalance issue. For the RAVDESS, accuracy is applied to better compare with other works~\cite{r1,r2,r3,r4}. %BS: the explanation comes below why you use different measures, but would be better to explain or forward reference at this point.

\sstitle{Implementation Details}
In our experiments, all audio recordings are re-sampled into $16$\,kHz. The batch size is $16$. 
For the upstream wav2vec 2.0 model fine-tuned on the RAVDESS, the fine-tuning procedure is optimised by an Adam optimiser with a learning rate of $3e-5$, and stopped after $20$ epochs. The feature encoder of wav2vec 2.0 is frozen. The outputs of the second last fully connected layer `FC1' of the wav2vec 2.0 model are averaged over the time frame, resulting in the embeddings $h$. The dimension of $h(x_i)$ is $256$, and we set the threshold $\epsilon$ as $0.99$. A lager max number of neighbors $n$ requires more computing resources on edge devices; therefore, we limit $n$ to under 10 considering the dataset size of the DEMoS. For ablation study, we choose 3, 6, and 9 for $n$. Similarly, we empirically chose 0.01, 0.1 and 1 for the multiplier $\alpha$.
%The maximum number of neighbours $n$ is set as $3$, $6$, and $9$ for experimental comparison. The multiplier $\alpha$ is experimented with $0.01$, $0.1$, $1$.

As for the downstream lightweight models development, log Mel spectrograms are extracted from the DEMoS audio samples as features. Firstly, we unify the audio length as the maximum one of all audio durations
%as $5.884$\,seconds 
%BS: why this value? Why not choose the max? Is it the mean?
by simply cutting extra signals and self-repeating shorter samples. To extract log Mel spectrograms, the sliding window, overlap, and Mel bins are set as $512$, $256$, and $64$ time frames, respectively. In this way, the extracted log Mel spectrograms' dimension is ($373$, $64$), where $373$ is on the time axis, and $64$ describes the number of Mel frequency bins. The model development is also optimised with an Adam optimiser with an initial learning rate of $1e-3$ and stopped after $50$ epochs. The learning rate decays by a factor of $0.9$ after every 5 epochs. As for the comparison with transfer learning, the same embeddings generated from the upstream wav2vec 2.0 model are added / maximised / averaged with the output of the `FC1' layer of the downstream lightweight models.
The code for this study has been made available at \url{https://github.com/glam-imperial/NSL-SER}.
%A weighted variant of cross-entropy is applied to better mitigate the class imbalance effect in the MSP-Podcast. Specifically, the weight given to each class is its corresponding reversed frequency.

\subsection{Wav2Vec2 fine-tuning}
\label{sec:w2v}

% When the encoder is non-trainable, the best validation UAR 55.80\% is achieved after 3 epoch; when the encoder is trainable, the best best validation UAR 53.76\% is obtained after 4 epoch. Therefore, on the train set and validation set, we set the epoch numbers as 3 and 4 for the trainable and non-trainable encoders, respectively. 

To ensure the efficacy of NSL in constructing the graphs, we compare the performance of the fine-tuned wav2vec 2.0 model with state-of-the-art (SOTA) models on the RAVDESS. Accuracy is used to ensure a relatively fair comparison. Moreover, the eight emotional classes are balanced besides \emph{neutral}, so accuracy is close to UAR. The SOTA models' accuracy scores on the RAVDESS test dataset are $76.08$\,\%~\cite{r1}, $78.8$\,\%~\cite{r2}, $93.2$\,\%~\cite{r3}, and $81.94$\,\%~\cite{r4}. Our fine-tuned wav2vec 2.0 model achieves $82.50$\,\% test accuracy, which is 
%comparable with the SOTA works. 
better than most works. Please note that the data split is different between SOTA's works and our work. Specifically, data split ratio, strategy, and considered emotional classes are not the same. %BS: What do you mean? Are these results comparable at all? Sounds like a killer...!

%As depicted in Table~\ref{tab:w2v}, the fine-tuned wav2vec2 model achieves comparable classification performance with SOTA works. Therefore, the wav2vec2 model is effective in emotion recognition and can be utilised to generate the embeddings in the neutral structured learning.

%\begin{table}[!h]
%\centering
%\footnotesize
%\vspace{-10pt}
%\caption{Classification performance [\%] comparison with the SOTA approaches.}
%\label{tab:w2v}
%\vspace{-10pt}
%\begin{tabular}{lll}
%\hline
% & Val Acc & Test Acc \\ \hline
%m1~\cite{r1} & -- & 76.08 \\
%m2~\cite{r2} & -- & 78.8 \\
%m3~\cite{r3} & -- & 93.2 \\
%m4~\cite{r4} & -- & 81.94 \\ \hline
%\textbf{Fine-tuned Wav2Vec2} & 65.00 & \textbf{82.50} \\ \hline
%\end{tabular}
%\vspace{-10pt}
%\end{table}

\begin{figure}[!h]
    \centering
    \begin{subfigure}[b]{0.85\linewidth}
    \centering
        \includegraphics[width=1\linewidth]{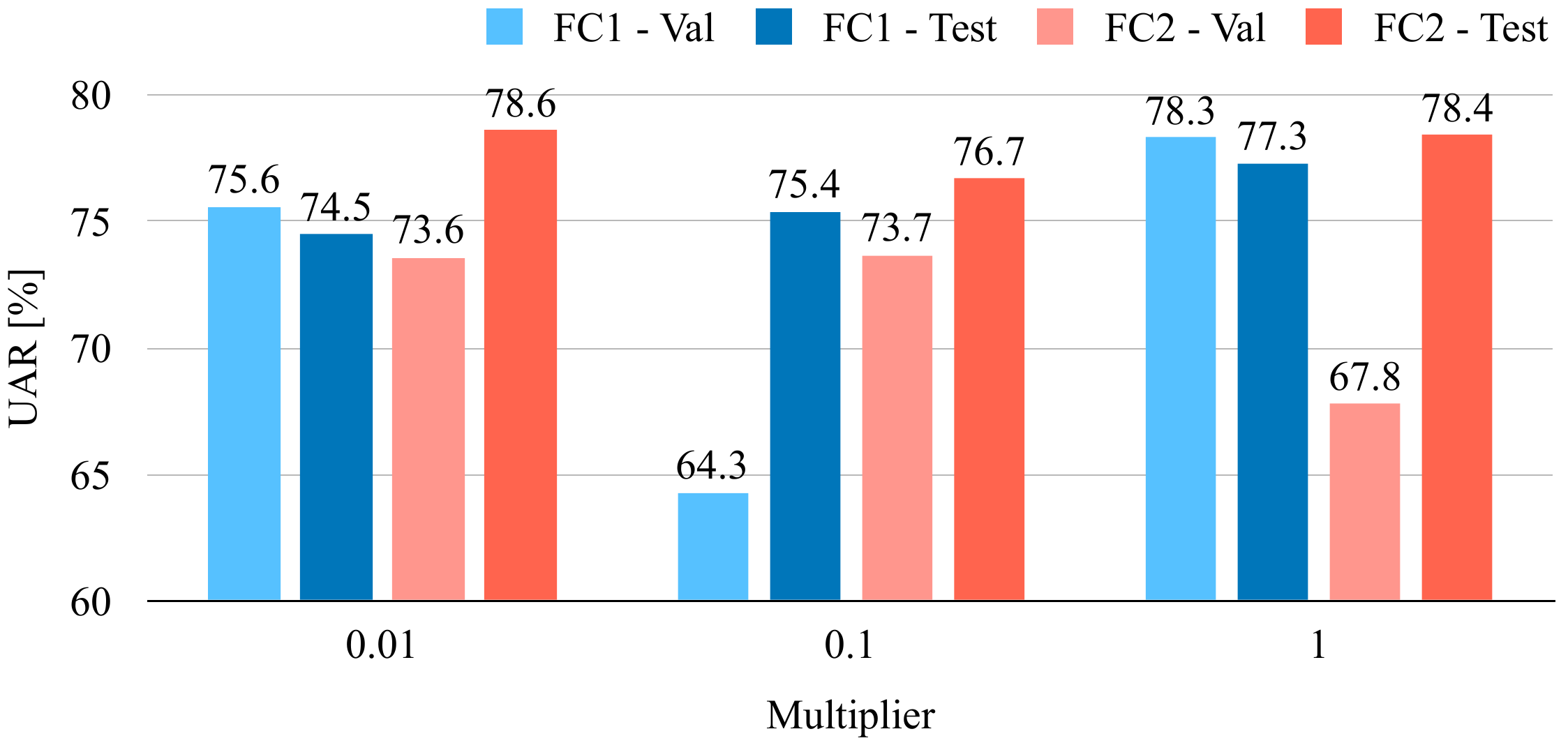}
        \vspace{-10pt}
        \caption{\# Neighbours = 3 }
        \label{fig: nsl3}
    \end{subfigure}
    \\
    \begin{subfigure}[b]{0.85\linewidth}
    \centering
        \includegraphics[width=1\linewidth]{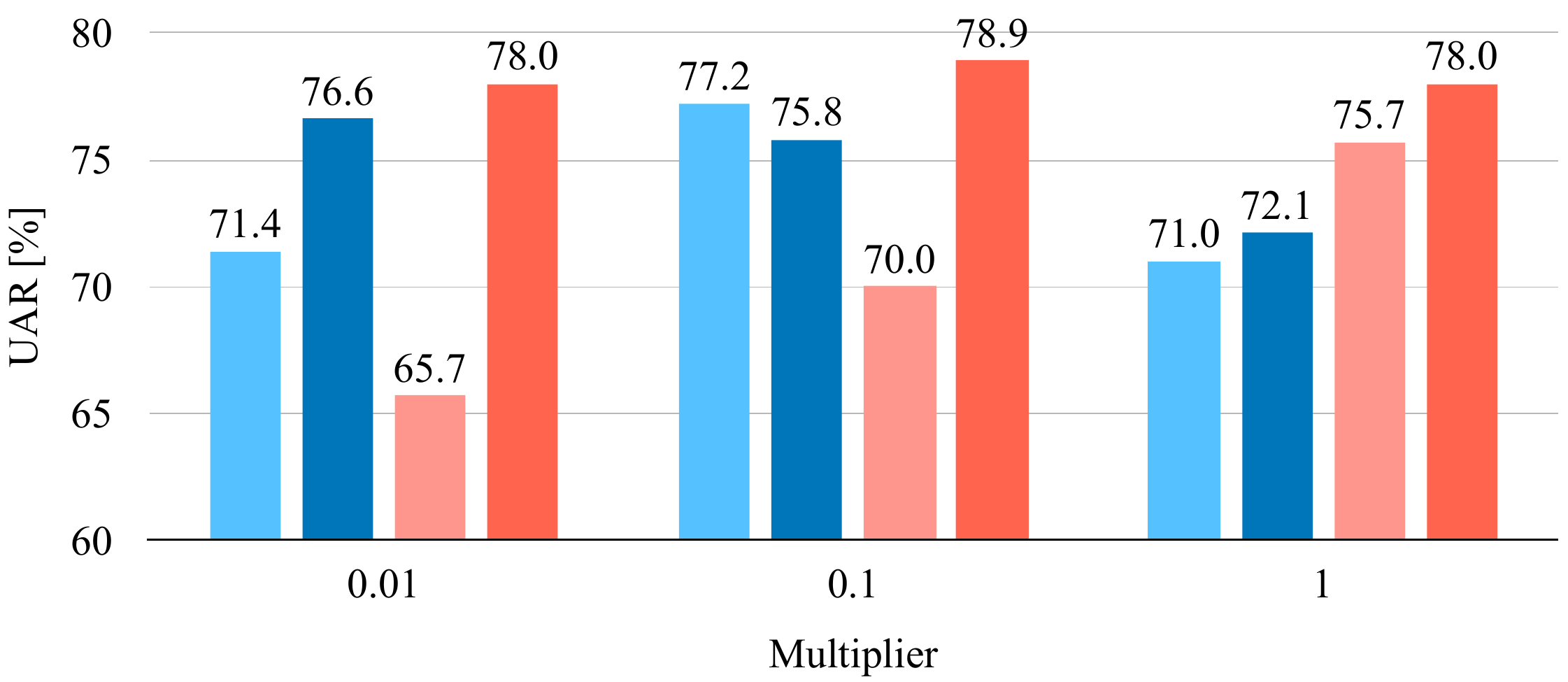}
        \vspace{-10pt}
        \caption{\# Neighbours = 6 }
        \label{fig: nsl6}
    \end{subfigure}
    \\
    \begin{subfigure}[c]{0.85\linewidth}
    \centering
        \includegraphics[width=1\linewidth]{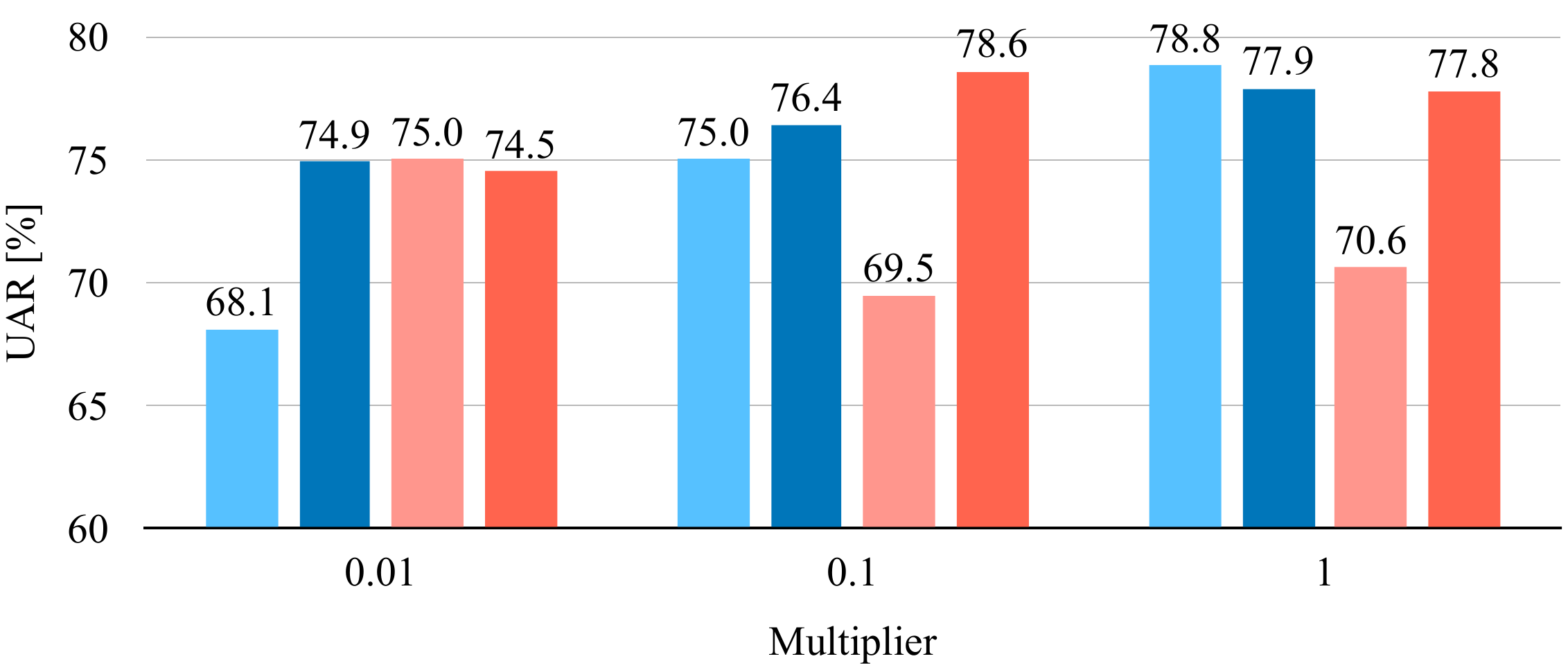}
        \vspace{-10pt}
        \caption{\# Neighbours = 9 }
        \label{fig: nsl9}
    \end{subfigure}
    \vspace{-5pt}
    \caption{Comparison of the performance (UAR [\%]) of the VGG-15 with NSL on the DEMoS validation and test datasets.}
    \vspace{-15pt}
    \label{fig:nsl-neighbor}
\end{figure}

\begin{table*}[!t]
%\footnotesize
\centering
\caption{Models' performances {[}Accuracy / UAR \,\%{]} on DEMoS validation and test datasets. $n$: maximum neighbours, $\alpha$: multiplier.}
%\resizebox{\columnwidth}{!}
\begin{adjustbox}{width=1.5\columnwidth,center}
\begin{tabular}
{l|r|r|P{1.5cm}|P{1.5cm}|P{1.5cm}|P{1.5cm}|P{1.5cm}|P{1.5cm}}
\toprule

& & &\multicolumn{2}{c|}{\textbf{VGG-15}} & \multicolumn{2}{c|}{\textbf{ResNet-9}} & \multicolumn{2}{c}{\textbf{CNN-6}}\\
\cline{4-9}

NN& \# $n$& $\alpha$ &\emph{Val} & \emph{Test} & \emph{Val} & \emph{Test} & \emph{Val} & \emph{Test}\\
\hline

Base Model & --& --& 71.0 / 66.4  & 84.0 / 77.2 & 79.4 / 73.9  & 81.7 / 74.6 & 68.4 / 65.3  & 80.7 / 73.1 \\
\hline

% \rowcolor{red}
Transfer Learning (Add) & --& --& 70.9 / 66.7 & 80.9 / \textbf{74.0} & 78.5 / 73.2  & 82.4 / \textbf{75.2} & 80.6 / 74.7 &  82.0 / \textbf{75.2} \\
\hline

% \rowcolor{red}
Transfer Learning (Max) & --& --& 76.3 / 70.1 & 75.1 / 68.5 & 80.3 / 74.9  & 81.1 / 74.5 & 74.2 / 67.8 & 76.0 / 70.0 \\
\hline

% \rowcolor{red}
Transfer Learning (Avg) & --& --& 70.0 / 63.8 & 74.3 / 66.4 & 82.0 / 76.8 & 82.5 / 75.1 & 80.5 / 74.8 & 81.6 / 74.0 \\
\hline

& 3 & 0.01 & 80.3 / 73.6 & 86.0 / 78.6 & 78.3 / 73.0 & 83.8 / \textbf{76.3} & 78.2 / 72.6 & 84.2 / \textbf{77.0} \\

NSL & 6 & 0.1 & 74.9 / 70.0  & 85.9 / \textbf{78.9} & 74.5 / 67.8  & 81.6 / 74.5 & 80.7 / 74.9  & 83.1 / 76.0 \\

& 9 & 0.1 & 75.4 / 69.5  & 86.3 / 78.6	& 74.8 / 68.5  & 83.0 / 76.2 & 77.8 / 73.0  & 84.1 / 76.9 \\
\bottomrule
\end{tabular}
\label{tab:res}
\end{adjustbox}
\end{table*}

\subsection{Results and Sensitivity Analysis}
\label{sec:sensiticity}
As shown in Figure~\ref{fig:nsl-neighbor}, we evaluate the performance of VGG-15 models in different NSL settings, \ie different maximum neighbours $n$ and two embedding layers (`FC1', `FC2') for building graphs.
Compared with the performance of base VGG-15 (validation UAR $66.4$\,\% , test UAR $77.2$\,\%), VGG-15 with NSL achieves better performance.
Specifically, when $n$ is 6, $\alpha$ is 0.1, and the embedding layer is `FC2', the test UAR obtains significant improvement ($p < 0.1$ in a one-tailed z-test). To further validate the effectiveness of the proposed NSL framework, we choose the best settings for $n$, $\alpha$, and embedding layer based on the test UAR indicated in Figure~\ref{fig:nsl-neighbor}. Specifically, by averaging and comparing the test UAR grouped by `FC1' and `FC2', we observe that `FC2' tends to obtain better performance, possibly due to more abstract representations learnt after `FC2'. Afterwards, for different $n$ (\ie 3, 6, 9), we find the best multipliers are 0.01, 0.1, and 0.1, respectively. With more neighbours, stronger structural similarities among inputs are obtained, and the neighbour loss may play a more crucial role during model training with larger multiplier values. It further validates the efficacy of the proposed NSL framework. The following experiments are conducted with lightweight models with these settings.

\begin{figure}
    \centering
    \vspace{-10pt}
    \includegraphics[width=0.9\linewidth]{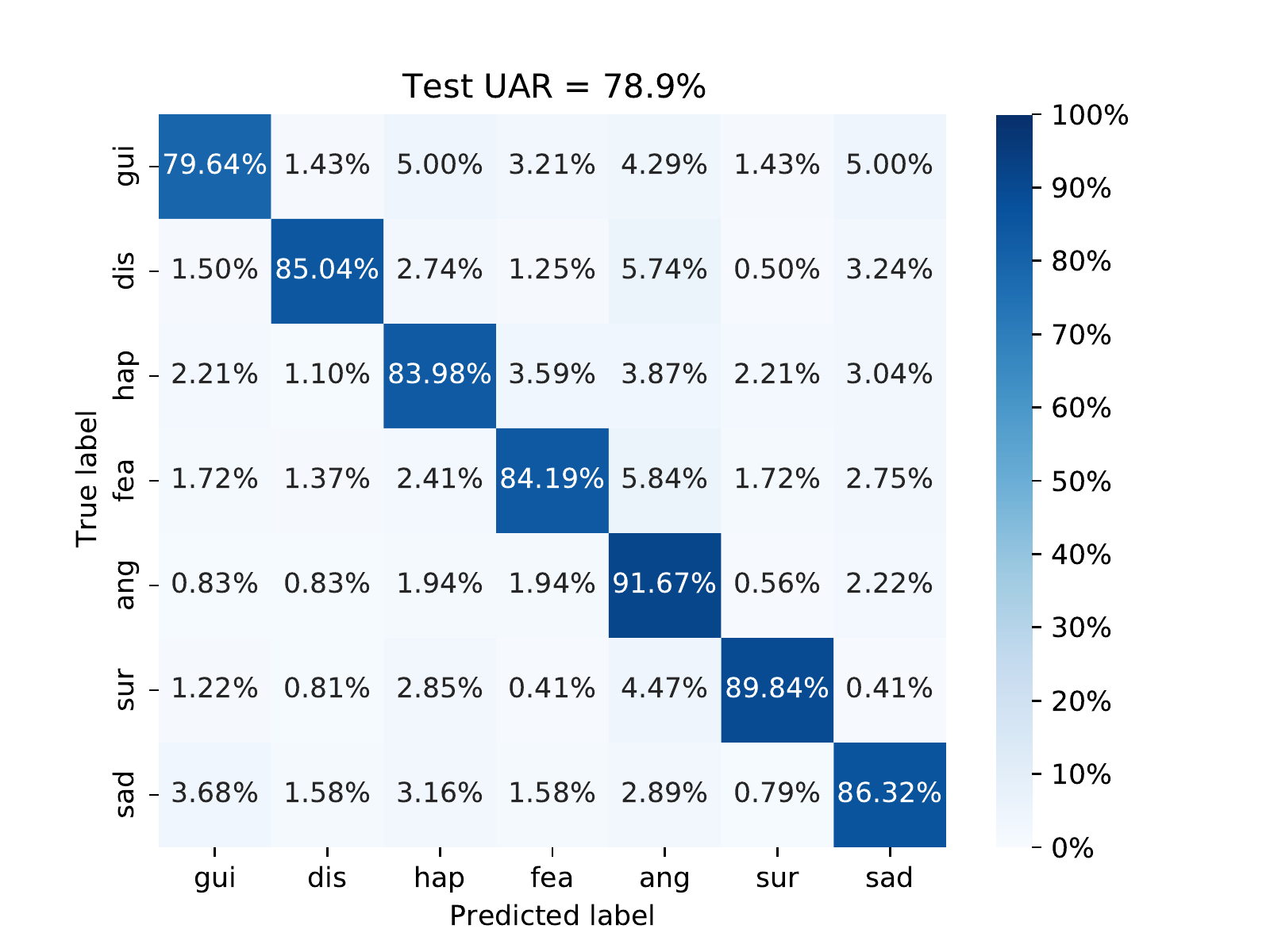}
    \caption{Confusion matrix for the best VGG-15 under NSL framework on the DEMoS test dataset. `gui': Guilt, `dis': Disgust, `hap': Happiness, `fea': Fear, `ang': Anger, `sur': Surprise, `sad': Sadness.}
    \vspace{-15pt}
    \label{fig:cf}
\end{figure}

Performances of VGG-15, ResNet-9, and CNN-6 on the DEMoS are presented in Table~\ref{tab:res}, where the `Base Model' refers to the model trained solely on log Mel spectrograms. On both accuracy and UAR, NSL models almost outperform base models. Specifically, when $n$ is 3 and $\alpha$ is $0.01$, test UARs of ResNet-9 (76.3\,\%) and CNN-6 (77.0\,\%) significantly improve compared to their corresponding base models ($p < 0.1$ and $p < 0.002$ in one-tailed z-tests, respectively). Notably, for CNN-6, NSL substantially enhances almost all validation and test UARs, indicating significantly better performance. These results demonstrate NSL can help improve simple CNNs' performance via knowledge transfer. Comparing with classic transfer learning strategies, NSL models also perform better. Specifically, test UARs of VGG-15 and CNN-6 are significantly better than those of transfer learning models ($p < 0.001$ and $p < 0.1$ in one-tailed z-tests, respectively). Figure ~\ref{fig:cf} shows the class-wise analysis of best model VGG-15 with NSL on the DEMoS test dataset.

\subsection{Discussion}
\label{sec:dicussison}
With graphs maintaining the structural similarities among input signals, NSL enables performance improvement of lightweight models. However, it should be noted that some of the results in Table~\ref{tab:res} are not stable, such as the validation UARs of ResNet-9 with NSL. This instability may be caused by the latent data difference between RAVDESS and DEMoS. Specifically, the DEMoS dataset contains speech in Italian, whereas samples in RAVDESS are in English; furthermore, the emotion classes of the two datasets are also different. When compared to the SOTA works on DEMoS~\cite{ren2020generating}, our work's performance is comparable, which can be attributed to data split differences. However, we would like to emphasise that this work focuses on the performance improvement of lightweight models with the help of knowledge transfer by the NSL framework. 

% \begin{table*}[htp]
% %\footnotesize
% \centering
% \caption{Models' performances {[}Accuracy / UAR \,\%{]} on DEMoS validation and test datasets.}
% %\resizebox{\columnwidth}{!}
% \begin{adjustbox}{width=1.95\columnwidth,center}
% \begin{tabular}
% {l|l|l|P{1.3cm}|P{1.3cm}|P{1.3cm}|P{1.3cm}|P{1.3cm}|P{1.3cm}}
% \toprule

% & & &\multicolumn{2}{c|}{\textbf{VGG-15}} & \multicolumn{2}{c|}{\textbf{ResNet-9}} & \multicolumn{2}{c|}{\textbf{CNN-6}}\\
% \cline{4-9}

% & & &\emph{Val} & \emph{Test} & \emph{Val} & \emph{Test} & \emph{Val} & \emph{Test}\\
% \hline

% Base Model Learning & & & 71.0 / 66.4  & 84.0 / 77.2 & 79.4 / 73.9  & 81.7 / 74.6 & 68.4 / 65.3  & 80.7 / 73.1 \\
% \hline

% & \# Neighbors & Multiplier & & &  & & & \\

% & 3 & 0.01 & 80.3 / 73.6 & 86.0 / 78.6 & 78.3 / 73.0 & 83.8 / \textbf{76.3} & 78.2 / 72.6 & 84.2 / \textbf{77.0} \\

% Neural Structured Learning & 6 & 0.1 & 74.9 / 70.0  & 85.9 / \textbf{78.9} & 74.5 / 67.8  & 81.6 / 74.5 & 80.7 / 74.9  & 83.1 / 76.0 \\

% & 9 & 0.1 & 75.4 / 69.5  & 86.3 / 78.6	& 74.8 / 68.5  & 83.0 / 76.2 & 77.8 / 73.0  & 84.1 / 76.9 \\
% \bottomrule
% \end{tabular}
% \label{tab:res}
% \end{adjustbox}
% \end{table*}

\section{Conclusions and Future Work}
\label{sec:conclusion}
This paper proposed a neural structured learning (NSL) framework to transfer the knowledge learnt from the related speech emotion dataset to the target one by maintaining the structural similarities defined in graphs. Specifically, a pre-trained upstream wav2vec 2.0 model was fine-tuned on the RAVDESS 
%BS: added:
emotional speech database
for graph construction on DEMoS; further experiments on the DEMoS 
%BS: added: 
emotional speech database
with downstream lightweight models validated the effectiveness of our NSL framework. 

In future work, a larger-scale source dataset (\eg CMU-MOSEI~\cite{zadeh2018multi}) can be applied as the source dataset and more target datasets can be also explored. Moreover, some domain adaptation strategies~\cite{MAO20171} can be explored to bridge the latent data difference between the source dataset and the target dataset. 
%BS: perhaps you can fill until page end. For example, adding acknowledgements.

% \vfill\pagebreak

% References should be produced using the bibtex program from suitable
% BiBTeX files (here: strings, refs, manuals). The IEEEbib.bst bibliography
% style file from IEEE produces unsorted bibliography list.
% -------------------------------------------------------------------------
\balance
\bibliographystyle{IEEEbib}
\bibliography{main}

\end{document}